%% file: arxiv_30dec2025.tex
\documentclass[12pt]{article}\usepackage{amsmath}
\usepackage{bm}

\usepackage{amsfonts}
\usepackage[onehalfspacing]{setspace}
\usepackage{geometry}
\usepackage{amsmath}
\usepackage{footmisc}
\usepackage{xcolor}
\usepackage{verbatim}
\usepackage[T1]{fontenc}
\usepackage[utf8]{inputenc}

\makeatletter
\renewcommand{\@fnsymbol}[1]{\ensuremath{%
   \ifcase#1\or *\or **\or {**}*\or
   \mathsection\or \mathparagraph\or \|\or \star\or
   \star\star\or {\star\star}\star \else\@ctrerr\fi}}
\makeatother
\usepackage{hyperref}
\usepackage{amsmath}
\usepackage{epstopdf}
\usepackage{amsfonts}
\usepackage{amssymb}
\usepackage{graphicx}
\usepackage{rotating}
\usepackage{lscape}
\usepackage{float}

\graphicspath{{T:/server/chiron/projects/female/figures/}}

\begin{document}

\title{Persistence of Gender Norms and Women Entrepreneurship\vspace*{-0.2cm}\thanks{\scriptsize{We thank Luis Aguiar, Anne Brenøe,Tom Brökel, Marita Freimane, Patricia Funk, Lorenz Küng, Raphael Parchet, Klaus Meyer, Peter Moser, Carmit Segal, Rainer Winkelmann, Josef Zweimüller and in particular Moritz Lubczyk for very valuable feedback. Special thanks to Anna Stünzi for her early input and comments, and to Manuel Bolz, Barbara Go{\l}\k{e}bska and Natalia Kasi\'{n}ska for excellent research assistance. The Swiss business register data was collected in collaboration with \url{www.chiron-services.ch}. We gratefully acknowledge funding from the Swiss National Science Foundation (``Spark'' grant CRSK-220770), Innosuisse, as well as the Carlsberg Foundation and the Mærsk McKinney Møller Foundation. We are thankful for the comments we received during presentations at Copenhagen Business School as well as the Universities of Düsseldorf, Lausanne, Lugano and Zurich. ©2025. Licensed under the Creative Commons Attribution-NonCommercial-NoDerivatives (CC BY-NC-ND) license.}}}
\author{Ulrich Kaiser\vspace*{-0.2cm}\thanks{{\scriptsize {University of Zurich; 
ulrich.kaiser@business.uzh.ch; Copenhagen Business School, ZEW -- Leibniz Centre for European Economic Research, Mannheim, and IZA Institute of Labor Economics, Bonn.}}} \quad Jos\'{e} Mata\vspace*{-0.2cm}\thanks{{\scriptsize {Mærsk McKinney Møller Professor of Entrepreneurship, Copenhagen Business School; jm.si@cbs.dk.}}} }

\maketitle\thispagestyle{empty}

\begin{abstract}
We examine whether gender norms --- proxied by the outcome of Switzerland’s 1981 public referendum on constitutional gender equality --- continue to shape local female startup activity today, despite substantial population changes over the past four decades. Using startup data for all Swiss municipalities from 2016 to 2023, we find that municipalities that historically expressed stronger support for gender equality have significantly higher present women-to-men startup ratios. The estimated elasticity of this ratio with respect to the share of ``yes'' votes in the 1981 referendum is 0.165. This finding is robust to controlling for a subsequent referendum on gender roles, a rich set of municipality-specific characteristics, and contemporary policy measures. The relationship between historical voting outcomes and current women’s entrepreneurship is stronger in municipalities with greater population stability --- measured by the share of residents born locally --- and in municipalities where residents are less likely to report a religious affiliation. While childcare spending is not statistically related to startup rates on its own, it is positively associated with the women-to-men startup ratio when interacted with historical gender norms, consistent with both formal and informal support mechanisms jointly shaping women’s entrepreneurial activity.
\end{abstract}

\newpage
\setcounter{page}{1}\section{Introduction}
Despite progress in gender equality in labor markets, female startup founders remain significantly underrepresented worldwide (Duflo, 2012). Limited participation of women in economic activity substantially constrains economic growth (Hsieh et al., 2019), and limited participation in entrepreneurship may have an even greater impact --- both because of entrepreneurship’s central role in growth and innovation (Akcigit and Kerr, 2018; Guzman and Stern, 2020) and because fewer female entrepreneurs can lead to reduced availability of products designed specifically for women (Koning et al., 2021).

Women’s underrepresentation in entrepreneurship may reflect enduring identity-based and social norms that shape career choices (Akerlof and Kranton, 2000). Yet the influence of gender norms on gendered economic behavior remains poorly understood, as such norms are rarely explicit and are often intertwined with political institutions (North, 1991).

We provide evidence on the role of gender norms in women’s entrepreneurship, measured as the ratio of women-to-men startups at the municipality level, by exploiting two features of Switzerland’s unique institutional and cultural setting. First, Switzerland has a strong federal structure in which 2{,}337 municipalities are organized into 26 cantons that retain extensive decision rights over most aspects of political and economic organization. Second, Switzerland has a strong system of direct democracy, in which citizens are regularly called to vote on a wide range of issues. Governments, parliaments, and political parties make recommendations, but policy outcomes are ultimately determined by citizens’ votes. We focus on the 1981 public referendum on enshrining the principle of equality between men and women in the Swiss Constitution (Janssen et al., 2016; Lalive and Stutzer, 2010; Palffy et al., 2023). This referendum is particularly informative because it directly expressed prevailing gender norms. Unlike most other referenda --- which typically have direct, often budgetary, implications, including later votes on gender-related issues --- the 1981 referendum had no immediate material consequences. While it laid the foundation for subsequent gender-equality reforms, its implementation was delayed by a five-year transition period, underscoring its normative rather than material character.

We isolate the effect of gender norms from that of formal institutions by controlling for canton fixed effects, which absorb much of the time-invariant political and economic institutional environment. Identification therefore relies on variation in gender norms across municipalities within the same canton, as expressed in the 1981 vote.\footnote{Gender norms varied widely across Swiss cantons, as reflected in votes on women’s suffrage at both the national and local levels. Shortly after nationwide women’s voting rights were rejected in a national referendum in 1959, three French-speaking cantons nonetheless granted women the right to vote locally. A subsequent national referendum approved women’s suffrage in 1971, but this decision did not immediately translate into local electoral practice with several cantons continuing to resist women’s suffrage. In 1990, the Federal Tribunal ordered the last remaining canton to grant women voting rights at the regional and local levels, ruling such exclusion unconstitutional in light of the 1981 referendum (Federal Tribunal 1990; https://tinyurl.com/7mya7m7p).}

The period following the 1981 referendum coincided with profound demographic and cultural changes. Between 1980 and 2020, Switzerland’s population grew by approximately 60\%, largely driven by immigration. By 2023, around 40\% of the population aged 15 and older had a migration background, and 32\% were first-generation immigrants. Over the same period, the share of residents reporting a non-national main language spoken at home increased from 6\% to 23\%, while the share of residents without a Christian religious affiliation, including atheists, rose from below 5\% to 47\%.

Despite these substantial changes, we find a persistent relationship between the expression of gender norms in 1981 and women’s relative entrepreneurial activity today. A one percentage point increase in the share of ``yes'' votes in the 1981 referendum is associated with a 0.165 percentage point increase in the women-to-men startup ratio. The magnitude of this elasticity is roughly twice that of the elasticity with respect to the share of Green Party voters in recent national elections, a party widely regarded as strongly supportive of gender equality in Switzerland. This relationship remains stable when controlling for a broad set of potential confounders, including canton fixed effects, contemporary political preferences, religion, language, and municipal characteristics such as childcare provision and women’s overall labor force participation. The results are also robust to controlling for a subsequent referendum on women’s role in society held in 2013.

Finally, we show that the persistence of gender norms is more strongly associated with contemporary female entrepreneurship in municipalities with greater population stability, in municipalities where a larger share of residents report no religious affiliation, and in municipalities that invested more heavily in childcare facilities. The latter association, however, is present only in municipalities that were relatively supportive of gender equality in 1981 --- formal daycare provision is not sufficient to increase women startups rates as it additionally requires informal social support at the local level.

\section{Data}
This section describes the data we use to link municipal voting behavior in 1981 to contemporary gender gaps in entrepreneurship: the business register, the 1981 voting data as well as additional information we use gathered from public sources.

\subsection{Business register data}
At the core of our empirical analysis are the 26 cantonal business registers, the Zefix registers, from which we downloaded the data via the API \url{https://tinyurl.com/4ar4h354}. Our data start on March 2, 2016 --- the first date when the business register data became available online --- and end on December 31, 2023.

The registers do not include gender identifiers but contain gendered terms such as Staatsbürgerin (female citizen in German), citoyenne (French), or cittadina (Italian), from which sex was identified. Cases that do not possess such explicit gender markers were classified using \url{www.gender-api.com}, a widely used platform for probabilistically associating names with gender (e.g., Aguiar, 2023). To handle ambiguous names (Andrea is predominantly male in Italian but female in German), we combine first names with country or region of origin. Cases in which the assigned probability is below 0.6 (fewer than one percent of all startups) were classified as “unknown” with respect to gender.

During our observation period, there were 346,356 unique new business registrations in our data. Of these, 60,493 (17.5\%) were founded exclusively by women, 215,928 (62.3\%) were founded exclusively by men, 50,532 (14.6\%) were gender-balanced (team startups with as many women as men), 3,293 (0.95\%) had a “female majority” (team startups with more women than men), and 12,838 (3.7\%) had a ``male majority''. We henceforth define ``women startups'' as firms that are ``dominated'' by women, that is, firms in which the number of women founders is greater than that of men. There are 63,786 (18.4\%) women-dominated firms in our data, compared to 228,766 (66\%) men-dominated startups. We aggregate the startup-level data to the municipality level, as we want to relate the prevalence of women’s entrepreneurship to the results of the vote held in 1981 on gender equality, for which we have municipal data.

\subsection{1981 voting data}
The 1981 referendum on the inclusion of an article establishing equality of rights between men and women in the Swiss Constitution was approved by 60.2\% of voters. Support varied substantially across municipalities, with approval rates ranging from 0\% to 100\% (unweighted municipal mean: 53\%).

The voting data were provided upon request by the Swiss Federal Statistical Office (FSO). The dataset originally covered 2,900 municipalities, of which 125 are missing; according to the FSO, this represents the most accurate data available. Due to subsequent municipal mergers, the number of Swiss municipalities had decreased to 2,136 by 2023. To ensure consistency over time, we mapped the 1981 municipality codes to their corresponding 2016–2023 codes using an API provided by the FSO (\url{https://tinyurl.com/26mv5u4v}).

					\begin{figure}[htbp]
					\caption{\textbf{Share of ``Yes'' Votes 1981}}
					\label{mapsofswitzerland}
					\begin{center}
						\includegraphics[width=.8\linewidth]{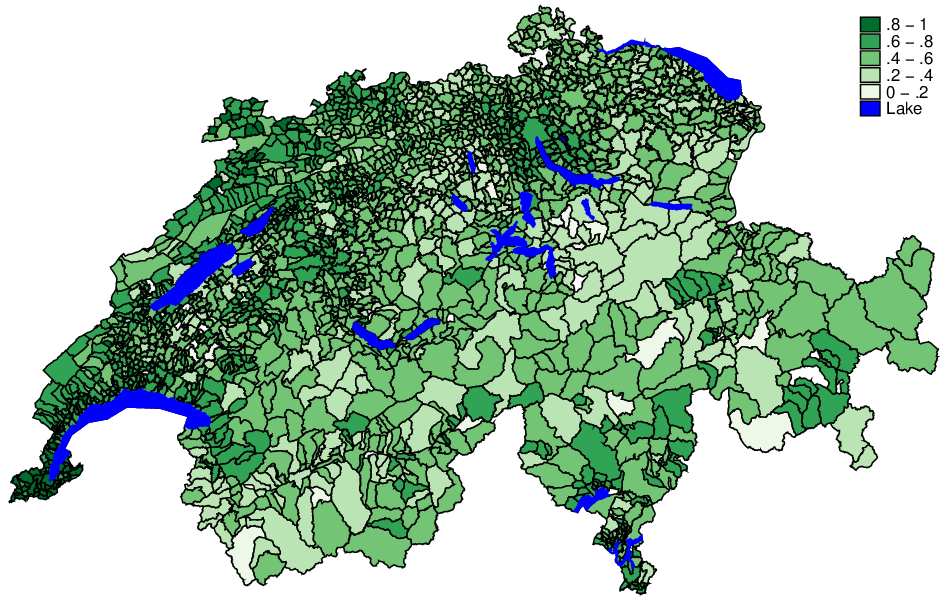}
					\end{center}
					\end{figure}	

					\begin{figure}[htbp]
					\caption{\textbf{Women startup rates}}
					\label{mapsofswitzerland}
					\begin{center}
						\includegraphics[width=.8\linewidth]{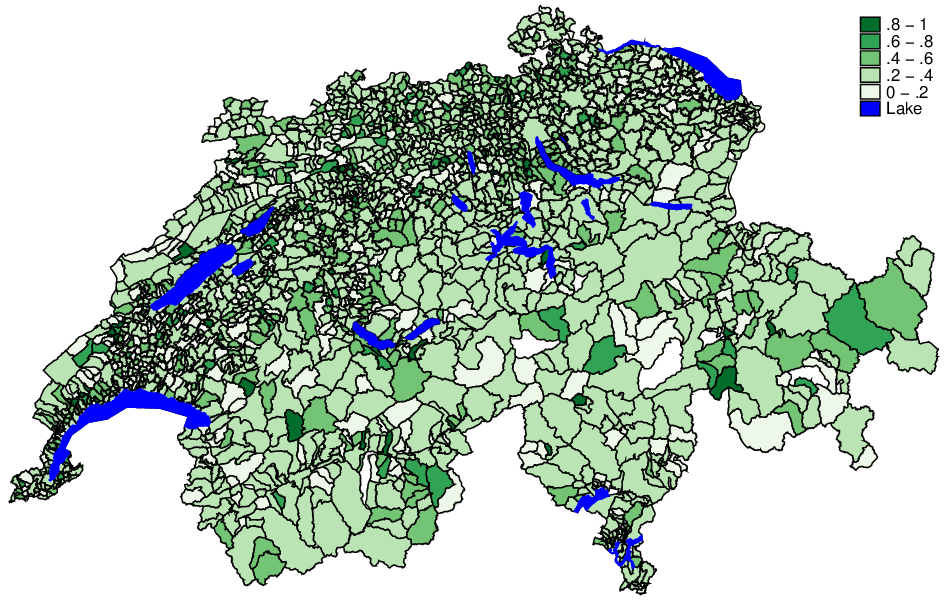}
					\end{center}
					\end{figure}	

\subsection{Empirical approach}
Our dependent variable is the ratio of women-dominated startups in municipality $i$ ($W_i$) relative to the number of men-dominated startups ($M_i$). This ratio allows us to focus on the determinants of women’s versus men’s startup activity while accounting for unobserved, municipality-specific factors associated with general entrepreneurial activity that are independent of gender. Our OLS estimating equation is $\frac{W_i}{M_i} = \alpha \text{YesShare}_{1981,i} + \bm{X}_i \bm{\beta}+\varepsilon_i$, where $\alpha$ is the parameter of primary interest and $\bm{\beta}$ captures how the women’s startup ratio varies with the vector of potential confounders $\bm{X}_i$. We cluster standard errors at the labor market region level, as municipalities within the same labor market region are likely to be exposed to common unobserved shocks, inducing correlation in their error terms. The 101 unique labor market regions in our data are defined based on regional commuting patterns by the FSO.\footnote{\url{https://tinyurl.com/yf4k455v}}

Clearly, $\alpha$ is causally identified if and only if the share of ``yes'' votes in 1981 is conditionally independent of the error term $\varepsilon_i$. That is, after controlling for the set of potential confounders $\bm{X}_i$, $\varepsilon_i$ and $\text{YesShare}_{1981,i}$ are independent.

\subsection{Specification}
We include five sets of potential confounders in our empirical specifications. First, we include canton fixed effects, which allow us to control for formal institutional differences across cantons. Identification therefore relies on within-canton variation between municipalities.

Second, we control for structural characteristics of municipalities, motivated by the well-documented urban--rural divide in Switzerland (e.g., Zumbrunn, 2024). Specifically, we include time-invariant dummy variables for urban and rural municipalities provided by the FSO.\footnote{\url{https://tinyurl.com/3naa4sy4}} In addition, we control for the number of firms, the share of women in the population, and the extent of women’s labor market participation. Municipalities with a higher share of working women are likely to offer relatively more dependent employment opportunities for women, which may affect women’s startup activity.

Third, we account for municipalities’ contemporary political leanings by including the voting shares of the major Swiss political parties in the 2015 national elections. Because these voting shares may capture general political preferences but may not fully reflect attitudes toward gender equality, we also include voting outcomes from the 2013 ``Family Initiative’’.' This initiative, which was rejected, proposed tax benefits for stay-at-home parents and would therefore have favored more ``traditional'' gender roles.

Fourth, women’s startup ratios are likely to be influenced by childcare provision. As a highly federalized country in which municipalities are primarily responsible for childcare provision, Switzerland lacks comprehensive municipality-level childcare data. However, the federal government subsidizes the establishment and expansion of childcare facilities, and corresponding data are available from the Federal Social Insurance Office.\footnote{\url{https://tinyurl.com/bpvseab2}} We include the sum of all childcare-related expenditures between 2003 and 2015, scaled by the number of children in daycare, preschool, primary, or secondary school (i.e., generally children under the age of 17).

Fifth, we account for long-lasting cultural influences that may shape gendered entrepreneurship, particularly those conveyed through religion and language. Following Guiso et al. (2003), we proxy cultural norms using religious denomination by including the shares of Catholics, adherents of other religions, and individuals with no religious affiliation, with Protestants as the reference group. In Switzerland, ethnic groups are commonly identified by language (Alesina et al., 2003); accordingly, we include the shares of households in which French, Italian, German (the reference category), or another language is spoken to account for ethnic differences.\footnote{Switzerland’s fourth official language, Rhaeto-Romanic, is spoken by less than 0.5\% of the Swiss population (\url{https://tinyurl.com/5n749zxb}) and hence included in the ``other'' language category, which is essentially comprised of foreign languages. }

Both the language and religion variables are drawn from the Swiss Census 2000, the most recent dataset available at the municipality level. Voting data for the 2013 ``Family Initiative'' are obtained from \url{https://swissvotes.ch}, a repository maintained by the University of Bern. All remaining explanatory variables are sourced from the FSO's data catalogue.\footnote{\url{https://tinyurl.com/5yh97tkv}} Online Appendix A provides direct links to the respective datasets.

\subsection{Descriptive statistics}
Figure 1 displays the results of the 1981 referendum and Figure 2 women’s startup ratios across municipalities. The voting map reveals a pronounced rural--urban divide: urban municipalities were --- and remain --- more supportive of gender equality than rural areas. There is also a clear linguistic divide, with German-speaking regions being less supportive of gender equality than the traditionally more progressive French-speaking regions (Erhardt and Haenni, 2022). While the 1981 referendum exhibits strong geographic patterns, no similarly clear geographic structure emerges for women’s startup ratios. Correspondingly, there is no straightforward visual relationship between the two maps in Figure 1, at least prior to accounting for heterogeneity, as further illustrated by the scatterplot of the 1981 vote share against women’s startup ratios in Online Appendix B.

Table 1 reports descriptive statistics for all variables except the 26 canton fixed effects. The municipality-level average women-to-men startup ratio is 0.32, with substantial variation across municipalities (the standard deviation exceeds 70\% of the mean). To put this ratio into perspective, Table 1 also reports the average number of women-dominated and men-dominated startups per municipality, although these variables are not used directly in the estimations. On average, municipalities host 3.8 women-dominated startups and 13.7 men-dominated startups. In 29 municipalities, there are no men-dominated startups.

			{\tiny\input{summary1.tex}}

Almost 21\% of municipalities are classified as urban and 31\% as rural, with the remaining half categorized as intermediate. The strongest political party in Switzerland, the SVP, adopts a conservative stance on women’s role in society and received an average municipal vote share of 33.7\% in the 2015 national elections. The Evangelical People’s Party shares the SVP’s conservative orientation but accounted for only 1.6\% of the vote. At the opposite end of the political spectrum, the progressive Green Party received 5.9\% of the vote. The conservative referendum launched by the SVP in 2013 obtained an average approval share of 46.4\%.

Across municipalities, women’s average labor market participation rate is 41.5\%. With respect to childcare provision, municipalities invested an average of CHF~1{,}960 per child in childcare facilities between 2003 and 2015 (equivalent to USD~1{,}960 given purchasing power parity in 2015), although 59\% of municipalities recorded no childcare expenditures over this period.\footnote{Investments in childcare, scaled by the number of children of compulsory school age, and the share of women in the workforce are only moderately correlated with both the 1981 and the 2013 referendum outcomes. In addition, the referendum results and the 2015 national election vote shares of the SVP and the Green Party are only weakly correlated, as are the 1981 referendum outcome and childcare provision. The highest pairwise correlations are between the 1981 referendum vote share and the voting shares of the SVP and the Green Party, with correlation coefficients of $-0.36$ and $0.20$, respectively.}

\subsection{Main results}
Table 2 presents our main estimation results. We sequentially introduce explanatory variables, starting with the baseline specification in column (1), which includes only the 1981 voting share. In this specification, the coefficient estimate of the share of 1981 ``yes'' votes is 0.051 and statistically insignificant ($p$-value $0.183$). Introducing canton fixed effects as well as urban and rural municipality dummies in column (2) doubles the coefficient estimate and renders it statistically significant ($p$-value $0.01$). The estimated coefficient on the share of ``yes'' votes remains remarkably stable as additional sets of controls are added. Our full specification, which includes the baseline controls from column~(2) as well as contemporary political preferences, women’s labor force participation, religion, and language, yields a point estimate of 0.10 ($p$-value $0.054$).

The set of contemporary policy-related variables --- namely national election outcomes and the 2013 ``Family Initiative,'' labeled ``Share yes votes 2013'' in the table --- materially affect the estimated coefficient on the 1981 vote share, and the result of the 2013 vote is both statistically and economically insignificant. Taken together, these results suggest that gender norms, as proxied by the outcomes of the 1981 referendum, are highly persistent, even after accounting for more recent votes on women’s roles in society and contemporary political preferences.

								\begin{table}[htp]
								\caption{Main OLS Estimation Results}
								\begin{center}
								{\small\input{main.tex}}
								\end{center}
								 \vspace*{-0.4cm}
								\parbox{15.5cm}{\scriptsize{\textbf{Notes:} Clustered standard errors in parentheses. *** p$<$0.01, ** p$<$0.05, * p$<$0.1.}}
								\end{table}

Turning to the remaining covariates, the logarithm of the number of active firms in a municipality in 2015 is negatively and statistically significantly related to the women startup ratio. Municipalities with a larger number of firms are likely to offer more and better-paid wage employment opportunities, reducing incentives for women to start businesses (Guzman and Stern, 2020; Moretti, 2012). Similarly, municipalities with higher female labor force participation have relatively fewer women-dominated startups, consistent with greater availability of attractive dependent employment opportunities. Daycare expenditures are not significantly correlated with women’s startup activity when included on their own --- a result to which we return below --- nor are the religion variables statistically significant.

In contrast, language differences are economically and statistically meaningful. Relative to German speakers (the omitted category), higher shares of French and Italian speakers, as well as speakers of other languages, are associated with significantly lower women startup ratios. Put differently, municipalities with a higher share of German speakers tend to have higher women startup ratios.

The estimated relationship between the 1981 referendum outcome and contemporary women’s startup activity is large relative to current policy measures and indicators of economic and societal conditions. The implied mean elasticity is 0.165 (standard error $0.085$, $p$-value $0.052$). This elasticity is almost six times larger in absolute value than that associated with the number of firms ($-0.029$, $p$-value $0.019$) and more than twice as large as the elasticity of the Green Party’s vote share in recent national elections. The elasticities associated with the shares of French and Italian speakers are approximately $-0.06$ and are estimated with high precision.

\subsection{Endogeneity}
The error term $\epsilon_i$ captures all factors that affect women’s startup ratio today but are not captured by the 1981 voting share or our controls. Potential omitted variables may induce correlation between the error term and the 1981 votes and create bias. To test robustness for omitted variable bias, we conduct a regression sensitivity analysis (Diegert et al., 2022), benchmarking potential unobserved confounders against the most statistically significant observed predictors: canton dummies (especially Thurgau, Zug, and Saint Gallen), the share of Italian speakers, the share of women, and the urban-area dummy. The associated breakdown points are 30.2\%, 98.9\%, 53.1\%, 99.7\%, 98.4\%, and 37.7\%, respectively. This implies an unobserved confounder would need an effect nearly as large as the Zug dummy, the share of Italian speakers, or the share of women for our results to break down; for the other variables, it would need to be one-third to one-half as large which seems economically implausible.

\subsection{Reinforcement and attenuation of gender norms }
We next examine three factors that may have caused the influence of 1981 gender norms to persist (or fade) between 1981 and the mid-2010s, thereby shaping contemporary women’s entrepreneurship rates: (i) population stability, measured by the share of residents born in the municipality; (ii) changes in religious beliefs, measured by the share of people with no religious affiliation; and (iii) gender-equality policies, proxied by childcare expenditures. For each factor, we interact the 1981 vote share with the respective measure. Table 3 reports our estimation results. All models include the full set of controls used in Table 2, column (5). None of the factors is statistically or economically associated with women’s entrepreneurship when entered on their own, but each proxy becomes statistically and economically significant once interacted with the 1981 vote share, as shown in Table 3.

								\begin{table}[h]
								\caption{OLS Estimation Results: Reinforcement and Attenuation of Gender Norms}
								\begin{center}
								{\footnotesize\input{mechanisms.tex}}
								\end{center}
								 \vspace*{-0.4cm}
								\parbox{15.5cm}{\scriptsize{\textbf{Notes:} Clustered standard errors in parentheses. *** p$<$0.01, ** p$<$0.05, * p$<$0.1.}}
								\end{table}

The first factor, population stability, is related to the extent of the intergenerational transmission of norms. Occupational choices reflect group norms and socially constructed identities (Akerlof and Kranton, 2000; Shephard and Hanynie, 2009). These norms are intergenerationally transmitted and internalized within the family (Booth et al., 2019; Hauge et al. 2023; Fernández and Fogli, 2009), and a high share of outsiders weakens this transmission channel. Since newcomers have not been socialized into local values, they are less likely to adopt or reinforce them, and even children who immigrated with their parents are more influenced by their family’s norms than by those of their community (Fernández, 2007; Fernández and Fogli, 2009). The influence of newcomers is complex: outsiders may assimilate and partially adopt local values, but assimilation is gradual (Blau, 2015), and the values of outsiders may also influence the values of the original population (Jessen et al. 2024; Bredtmann et al. 2020). Both channels weaken the transmission of traditional values compared to a population not subject to outside influences. Column (1) reports our estimates for the interaction between the 1981 ``yes'' voting share and the 2015 share of residents born in the municipality. The coefficient is positive and statistically significant. At lower ``yes'' shares, a higher locally born share is associated with lower women’s startup rates. The share of locally born residents ranges from 0\% to 85\% with a mean of 36.9\%; the marginal effect of the share of residents born in the same municipality is negative when the 1981 ``yes'' share is below 0.72, which is true for 90\% of municipalities, and is not statistically significant above that threshold. Thus, where more residents are locally born, women’s entrepreneurial activity rate is lower, and this effect is stronger where the shares of ``yes'' votes is lower.

The second factor builds on the idea that gender norms reflect religiosity, or the absence thereof. The historical distribution of traditional Swiss religions is correlated with the outcome of the 1981 referendum. Municipalities that were predominantly Catholic in the mid-nineteenth century voted 3--4 percentage points less than predominantly Protestant municipalities in favor of gender equality in 1981.\footnote{Appendix C reports OLS regressions of the 1981 vote on historical (1850-1860) religious affiliations.} This pattern points to a persistent influence of religion on gender norms and is consistent with evidence showing that Protestantism’s (Lutheranism’s) emphasis on educating both boys and girls fosters attitudes supportive of gender equality (Becker et al. 2024).

Since 1980, attitudes toward religion have shifted markedly. The share of residents reporting no religious affiliation increased from 7.5\% in 1980 to 35.6\% in 2023. Religiosity of any type is associated with more patriarchal attitudes and greater gender inequality (Guiso et al. 2003; Seguino, 2011), while non-religious individuals tend to hold more gender-egalitarian views than Catholics, Protestants, or Muslims (Schnabel, 2016). The growth of the non-religious population has been particularly pronounced since 2000 --- the last year for which municipality-level religious data are available --- when its share stood at 11.4\%. As a result, the municipal distribution of non-religious residents in 2000 is likely to be a poor proxy for the corresponding distribution in 2015.

To address this limitation, we estimate the municipality-specific share of residents with no religious affiliation in 2015 using observed shares from 1980, 1990, and 2000, along with additional covariates. We construct a municipality-year panel and predict the 2015 share using linear time trends interacted with canton fixed effects, as well as controls for the share of women in the municipality, the Green Party vote share, the Evangelical People’s Party vote share, an urban municipality dummy, the share of residents born in the municipality, and the shares of households speaking French, Italian, or another language, with German as the reference category.\footnote{We bootstrap standard errors in all regressions that use the estimated 2015 share of non-religious residents to account for the ``generated regressors'' problem, whereby the variance--covariance matrix is no longer block-diagonal (Wooldridge, 2010, Ch.~6). We use 10{,}000 bootstrap replications. In specifications that include the predicted 2015 value, we exclude the observed 2000 non-religious share, as the correlation between the two is 0.71.}

The interaction between the 1981 ``yes'' vote share and the predicted 2015 share of residents with no religious affiliation is negative and highly statistically significant. This indicates that higher levels of non-religiousness are associated with greater women’s startup activity, particularly in municipalities with low support for gender equality in 1981. Consistent with the interpretation that a large non-religious population weakens traditional norms, the marginal effect of non-religiousness remains positive until the 1981 ``yes'' vote share reaches 81\%, a threshold exceeded by fewer than 3\% of municipalities. Accordingly, non-religiousness is positively associated with women’s startup rates, with stronger effects in locations where historical gender norms were more conservative.

The final dimension we consider is the extent to which spending on childcare facilities reinforces the association between 1981 gender norms and contemporary women’s entrepreneurship. Municipality-level investment in childcare facilities is not, by itself, significantly correlated with startup rates, as shown in Table~2. However, when we interact the 1981 ``yes'' vote share with childcare spending per child, both the main-effect and interaction coefficients are statistically significant, with a negative coefficient on the main effect and a positive coefficient on the interaction.

The tipping point occurs at a ``yes'' vote share of 39.6\%. About 17\% of municipalities fall below this threshold, for which the marginal effect is negative, while 83\% lie above it. Between 35\% and 48\% ``yes'' vote shares --- 599 municipalities or 25.6\% --- the marginal effect is not statistically different from zero.

Access to childcare generally coincides with higher women’s labor-force participation, with larger effects for disadvantaged women (Albanesi et al. 2023; Dahl and Løken, 2024). Our results show that where preferences for gender equality were weak, formal childcare expansion is associated with lower women’s startup rates, while it is associated with higher startup rates where preferences for gender equality were strong. This pattern is consistent with formal childcare provision enabling mothers to work, while entrepreneurial activity additionally requires informal childcare support, which is more likely in municipalities that historically hadstronger preferences for gender equality, given that entrepreneurship involves less predictable work schedules (Delecourt and Fitzpatrick, 2021).\footnote{Childcare investments (formal support) and the 1981 ``yes'' vote share are only weakly correlated, with a correlation coefficient of 0.02.}

Figure~3 illustrates how the historic ``yes'' vote share interacts with childcare investments. We plot the two extreme ``yes'' vote shares, 0 and 1, both observed in the data, across the full range of childcare expenditures per 1{,}000 children. With low social support, higher childcare spending is associated with a decline in women’s startup ratios --- from a predicted 0.28 with no spending to 0.12 at the maximum. For an intermediate ``yes'' vote share of 0.5, startup rates change little with spending.

Municipalities with a ``yes'' vote share of 1 have the highest women’s startup rates, and increases in childcare expenditures are associated with the largest gains; predicted ratios range from 0.33 to 0.58. At low spending levels, differences between low- and high-support municipalities are small and not statistically significant, as confidence bands overlap. As expenditure levels increase, the differences between low- and high-support municipalities widen; at the highest spending level, the female-to-male startup ratio rises nearly fivefold as the ``yes'' vote share increases from 0\% to 100\%.

							\begin{figure}[htp]
								\vspace{-.1cm}
								\caption{\textbf{Predicted Relationship Between Startup Rates as well as Daycare Expenditures and the Share of Yes Votes in 1981}}
								\begin{center}
								\includegraphics[scale=1]{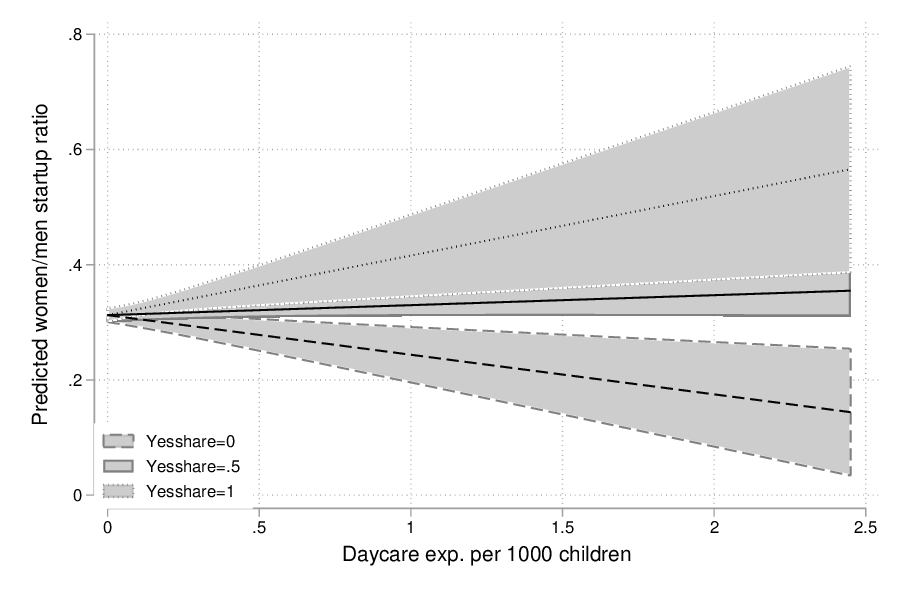}
								\label{overall}\\
								\end{center}
							\end{figure}

The final column of Table 5 displays the results of a specification where all interactions are simultaneously considered. While point estimates hardly change, estimation precision of the terms involving the 1981 voting shares of course decrease substantially. All combinations of all interactions are, however, still statistically significant. 

\section{Conclusions}
“Culture and social norms tend to evolve slowly, sometimes persisting over centuries” (Bisin and Verdier, 2001), but change can be swifter for issues that become central to society --- occurring within a generation or even less (Bertrand, 2011). However, despite the prominence of gender equality in Western societies, the gender values expressed in Switzerland’s 1981 referendum continue to be associated with women’s entrepreneurship more than four decades later. Non-religious attitudes weaken the hold of traditional norms, population stability reinforces it, and childcare spending can foster women’s entrepreneurship --- but only in communities that endorse gender equality in 1981 already. Formal policy measures alone may therefore be insufficient, as their effectiveness is mediated by culture and social support: supporting women’s engagement in entrepreneurship ultimately requires a supportive community.

\clearpage
\noindent {\bf\large Appendix A --- data sources}
\vspace*{-.75cm}
\begin{center}
\begin{table}[h]						
{\small
\begin{tabular}{lccl} \hline							
\hline							
Variable	&	Year	&	Level	&	Source	\\
\hline							
				\multicolumn{4}{l}{\bf Dependent variable} \\
				\# women and men startups & 2016-2023 & firm & \url{www.zefix.ch} \\
				\multicolumn{4}{l}{\bf Key explanatory variable} \\
				Share yes votes & 1981 & municipality & requested from BfS \\
				\multicolumn{4}{l}{\bf Additional explanatory variables related to:} \\
				Party strengths at national elections & 2015 & municipality & \url{https://tinyurl.com/4thbvj6p} \\
				Religion & 2000 & municipality & \url{https://tinyurl.com/96cc7dvf} \\
				Language & 2000 & municipality & \url{https://tinyurl.com/he47mryj} \\
				Municipality type & 2012 & municipality & \url{https://tinyurl.com/4svsscay} \\
				Place of birth & 2000 & municipality & \url{https://tinyurl.com/2rkuethh} \\
				Workforce participation & 2000 & municipality & \url{https://tinyurl.com/yc2xcrrk} \\
				Spendings on kindergarden children & 2001-2022 & municipality & \url{https://tinyurl.com/3axckjw8} \\
				Spendings on school children & 2001-2022 & municipality & \url{https://tinyurl.com/3axckjw8} \\
				\hline 
\end{tabular}}							
\end{table}							
\end{center}

				\noindent {\bf\large Appendix B --- Scatterplot 1981 Vote and Women Startup Ratios}\\
							\begin{figure}[htp]
								\vspace{-.3cm}
								\begin{center}
								\includegraphics[scale=.6]{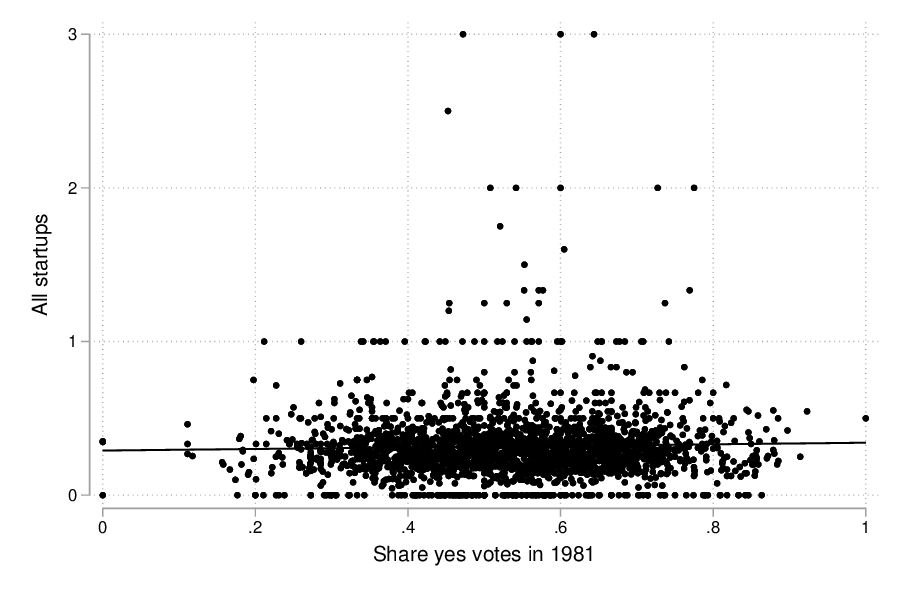}
								\label{overall}\\
								\end{center}
							\end{figure}

				\clearpage
				\noindent {\bf\large Appendix C --- 2015 Religion Prediction}\\
To predict the share of residents in a municipality who do not attain to any faith, we make use of the panel structure of our data. We regress the natural logarithm of share of non-religious residents on a linear and a quadratic time trend, the share of resisdents who originate from the same municipality to account for municipality-specific persistence in religion, the share of women in the municipality to account for women being on average more likely to practice religion (Stolz and Senn, 2022), the share of votes for the Green party to account for the progressiveness of the municipality, dummies for urban and rural municipalities to account residents of rural areas to be on average more religious (FSSPX News 2024), the share of children in a municipality to account for the age structure (Swissinfo, 2023), a set of canton dummies and the shares of speakers of French, Italian and languages other than the two and German.

The corresponding adjusted $R^2$ is 0.612 with a Root Mean Squared Error of 0.723. Given the dependent variable is in natural logarithms with a mean of -3.37, this implies that our prediction will be about $±2$ percentage points off on average, suggesting a good model fit.
We also tried a more flexible specification where all explanatory variables were interact with time trends which produced almost the exact same statistics. The predictions of the two models are correlated with a correlation coefficient of .94.

\clearpage

							\begin{table}[h]
							\begin{center}
							{\footnotesize\input{predict.tex}}
							\end{center}
							\parbox{15.5cm}{\scriptsize{\textbf{Notes:} Robust standard errors in parentheses. *** p$<$0.01, ** p$<$0.05, * p$<$0.1. The specification in column (2) additionally includes interactions of the canton dummies with a lienar time trend.}}									
							\end{table}

\end{document}

%% file: summary1.tex
\begin{table}[htbp]\centering \caption{Summary statistics \label{sumstat}}
\begin{tabular}{l c c  c}\hline\hline
\multicolumn{1}{c}{\textbf{Variable}} & \textbf{Mean}
 & \textbf{Std. Dev.} & \textbf{N}\\ \hline
Ratio of women to men startups & 0.318 & 0.231  & 2337\\
Share yes votes in 1981 & 0.535 & 0.143  & 2337\\
\# of firms in 2015 & 273.297 & 1119.87  & 2305\\
Daycare exp./1000 children & 0.196 & 0.996  & 2337\\
Share women in workforce & 0.415 & 0.07  & 2337\\
Share women residents & 0.497 & 0.016  & 2337\\
Urban municipality (d) & 0.205 & 0.404  & 2337\\
Intermediate municipality (d) & 0.484 & 0.5  & 2337\\
Rural municipality (d) & 0.311 & 0.463  & 2337\\
Voting share Swiss People's Party & 0.337 & 0.139  & 2337\\
Voting share Social Democrats & 0.152 & 0.07  & 2337\\
Voting share Liberal Democrats & 0.162 & 0.096  & 2337\\
Voting share Center & 0.182 & 0.139  & 2337\\
Voting share Greens & 0.059 & 0.044  & 2337\\
Voting share Green Liberals & 0.036 & 0.027  & 2337\\
Voting share Evangelical People's Party & 0.016 & 0.02  & 2337\\
Voting share Federal Democratic Union & 0.013 & 0.022  & 2337\\
Voting share other parties & 0.042 & 0.071  & 2337\\
Share yes votes 2013 & 0.464 & 0.086  & 2337\\
Share German speakers & 0.612 & 0.42  & 2337\\
Share French speakers & 0.268 & 0.4  & 2337\\
Share Italian speakers & 0.069 & 0.208  & 2337\\
Share other languages & 0.051 & 0.078  & 2337\\
Share protestants & 0.395 & 0.263  & 2337\\
Share catholics & 0.419 & 0.28  & 2337\\
Share other religions & 0.096 & 0.057  & 2337\\
Share no religion & 0.09 & 0.054  & 2337\\
Number of women startups in municipality & 3.77 & 16.887  & 2337\\
Number of men startups in municipality & 13.745 & 58.944  & 2337\\
\hline\end{tabular}
\end{table}

%% file: main.tex
\begin{tabular}{lccccc} \hline
 & (1) & (2) & (3) & (4) & (5) \\
VARIABLES & No controls & Cantons & Municipality & Politics & All \\ \hline
 &  &  &  &  &  \\
Share yes votes in 1981 & 0.051 & 0.113*** & 0.122*** & 0.111** & 0.098** \\
 & (0.183) & (0.010) & (0.012) & (0.019) & (0.054) \\
Urban municipality (d) &  & -0.040*** & -0.032*** & -0.030*** & -0.027*** \\
 &  & (0.000) & (0.003) & (0.002) & (0.006) \\
Rural municipality (d) &  & 0.005 & 0.012 & 0.013 & 0.015 \\
 &  & (0.684) & (0.368) & (0.349) & (0.286) \\
ln(\# firms) &  &  & -0.010** & -0.010*** & -0.009** \\
 &  &  & (0.020) & (0.012) & (0.021) \\
Daycare exp./1000 children &  &  & 0.015 & 0.014 & 0.015 \\
 &  &  & (0.306) & (0.306) & (0.280) \\
Share women in workforce &  &  & -0.083 & -0.086 & -0.110 \\
 &  &  & (0.463) & (0.443) & (0.333) \\
Share women residents &  &  & 1.337*** & 1.247*** & 1.145*** \\
 &  &  & (0.001) & (0.004) & (0.008) \\
Share yes votes 2013 &  &  &  & 0.001 & 0.013 \\
 &  &  &  & (0.996) & (0.924) \\
Share catholics &  &  &  &  & 0.025 \\
 &  &  &  &  & (0.638) \\
Share other religions &  &  &  &  & 0.047 \\
 &  &  &  &  & (0.663) \\
Share no religion &  &  &  &  & 0.316* \\
 &  &  &  &  & (0.058) \\
Share French speakers &  &  &  &  & -0.072** \\
 &  &  &  &  & (0.037) \\
Share Italian speakers &  &  &  &  & -0.259*** \\
 &  &  &  &  & (0.001) \\
Share other languages &  &  &  &  & -0.193* \\
 &  &  &  &  & (0.075) \\
 &  &  &  &  &  \\
Observations & 2,337 & 2,337 & 2,337 & 2,337 & 2,337 \\
R-squared & 0.001 & 0.025 & 0.040 & 0.047 & 0.056 \\
1981 vote & yes & yes & yes & yes & yes \\
Canton fixed effects & no & no & yes & yes & yes \\
Municipality characteristics & no & no & yes & yes & yes \\
Politics & no & no & no & yes & yes \\
Religion & no & no & no & no & yes \\
 Language & no & no & no & no & yes \\ \hline
\end{tabular}

%% file: mechanisms.tex
\begin{tabular}{lccccc} \hline
 & (1) & (2) & (3) & (4) & (5) \\
VARIABLES & Base & Social stability & Religion & Policy & All \\ \hline
 &  &  &  &  &  \\
Share yes votes in 1981 & 0.104** & -0.087 & 0.352*** & 0.064 & 0.222 \\
 & (0.042) & (0.386) & (0.001) & (0.214) & (0.256) \\
Yesshare * share same municipality &  & 0.422* &  &  & 0.099 \\
 &  & (0.070) &  &  & (0.706) \\
Share same municipality & -0.112 & -0.305** &  &  & -0.103 \\
 & (0.269) & (0.047) &  &  & (0.519) \\
Urban municipality (d) & -0.038* & -0.027*** & -0.016 & -0.027*** & -0.023 \\
 & (0.070) & (0.004) & (0.287) & (0.006) & (0.273) \\
Rural municipality (d) & 0.011 & 0.014 & 0.020 & 0.015 & 0.017 \\
 & (0.523) & (0.309) & (0.195) & (0.274) & (0.326) \\
ln(\# firms) & -0.010** & -0.010** & -0.009** & -0.010*** & -0.010*** \\
 & (0.021) & (0.018) & (0.020) & (0.008) & (0.009) \\
Daycare exp./1000 children & 0.015 & 0.015 & 0.015 & -0.065*** & -0.063 \\
 & (0.356) & (0.265) & (0.323) & (0.003) & (0.210) \\
Share women in workforce & -0.225 & -0.213 & -0.191 & -0.113 & -0.222 \\
 & (0.159) & (0.201) & (0.124) & (0.317) & (0.121) \\
Share women residents & 1.123*** & 1.152*** & 1.199*** & 1.186*** & 1.214*** \\
 & (0.014) & (0.008) & (0.006) & (0.007) & (0.005) \\
Share yes votes 2013 & 0.014 & 0.039 & -0.003 & -0.008 & -0.011 \\
 & (0.923) & (0.779) & (0.981) & (0.955) & (0.939) \\
Share catholics & 0.012 & 0.029 & 0.019 & 0.024 & 0.019 \\
 & (0.828) & (0.576) & (0.739) & (0.647) & (0.740) \\
Share other religions & 0.011 & 0.033 & 0.015 & 0.034 & -0.000 \\
 & (0.920) & (0.747) & (0.877) & (0.754) & (1.000) \\
Share no religion &  & 0.302* &  & 0.284* &  \\
 &  & (0.074) &  & (0.081) &  \\
Share French speakers & -0.060 & -0.074** & -0.091** & -0.070** & -0.079* \\
 & (0.201) & (0.032) & (0.032) & (0.035) & (0.070) \\
Share Italian speakers & -0.255** & -0.259*** & -0.228*** & -0.262*** & -0.242*** \\
 & (0.028) & (0.001) & (0.003) & (0.001) & (0.004) \\
Share other languages & -0.178 & -0.182* & -0.163* & -0.186* & -0.159* \\
 & (0.133) & (0.086) & (0.068) & (0.084) & (0.076) \\
Share no religion 2015 & -0.117 &  & 1.458*** &  & 1.038 \\
 & (0.822) &  & (0.011) &  & (0.167) \\
Yesshare * share no religion 2015 &  &  & -1.811*** &  & -1.429* \\
 &  &  & (0.006) &  & (0.073) \\
Yesshare * daycare expenditures &  &  &  & 0.165*** & 0.161 \\
 &  &  &  & (0.006) & (0.132) \\
 &  &  &  &  &  \\
Observations & 2,337 & 2,337 & 2,337 & 2,337 & 2,337 \\
R-squared & 0.056 & 0.058 & 0.058 & 0.065 & 0.066 \\
1981 vote & yes & yes & yes & yes & yes \\
Canton fixed effects & yes & yes & yes & yes & yes \\
Municipality characteristics & yes & yes & yes & yes & yes \\
Politics & yes & yes & yes & yes & yes \\
Religion & yes & yes & yes & yes & yes \\
Language & yes & yes & yes & yes & yes \\
 Interactions & yes & yes & yes & yes & yes \\ \hline
\end{tabular}

%% file: predict.tex
\begin{tabular}{lcc} \hline
 & (1) & (2) \\
VARIABLES & Restr. prediction & Flex. prediction \\ \hline
 &  &  \\
Time trend & -0.003 & -0.067* \\
 & (0.770) & (0.081) \\
Time trend squared & -0.001*** & -0.001*** \\
 & (0.000) & (0.000) \\
Share same municipality & -1.707*** & -0.797*** \\
 & (0.000) & (0.004) \\
Share women in workforce & 3.795*** & -1.702 \\
 & (0.000) & (0.484) \\
Urban municipality (d) & -0.023 & -0.300*** \\
 & (0.368) & (0.003) \\
Rural municipality (d) & -0.187*** & -0.109 \\
 & (0.000) & (0.199) \\
Share French speakers & 0.805*** & 0.588*** \\
 & (0.000) & (0.003) \\
Share Italian speakers & 0.465*** & -1.266*** \\
 & (0.000) & (0.004) \\
Share other languages & 0.250** & 0.024 \\
 & (0.040) & (0.961) \\
Share Evangelical People's Party & -1.098* & -1.974 \\
 & (0.082) & (0.421) \\
Share Green Party & 4.056*** & 3.601*** \\
 & (0.000) & (0.005) \\
Share children in population & -3.579*** & -0.051 \\
 & (0.000) & (0.986) \\
Share same municipality * time trend &  & -0.029*** \\
 &  & (0.001) \\
Share women in workforce * time trend &  & 0.178** \\
 &  & (0.019) \\
Urban municipality (d) * time trend &  & 0.009*** \\
 &  & (0.005) \\
Rural municipality (d) * time trend &  & -0.003 \\
 &  & (0.342) \\
Share French speakers * time trend &  & 0.007 \\
 &  & (0.222) \\
Share Italian speakers * time trend &  & 0.053*** \\
 &  & (0.000) \\
Share other languages * time trend &  & 0.009 \\
 &  & (0.558) \\
Share Evangelical People's Party * time trend &  & 0.028 \\
 &  & (0.713) \\
Share Green Party * time trend &  & 0.014 \\
 &  & (0.726) \\
Share children in population * time trend &  & -0.115 \\
 &  & (0.202) \\
 &  &  \\
Observations & 7,011 & 7,011 \\
R-squared & 0.641 & 0.651 \\
Time interactions & no & yes \\
 Canton dummies & yes & yes \\ \hline
\end{tabular}